\documentclass[english,aps,prstper,reprint,showpacs,longbibliography,superscriptaddress,nofootinbib]{revtex4-2}   

\usepackage[T1]{fontenc}	
\usepackage[latin9]{inputenc}	
\usepackage{geometry}		
\usepackage{graphicx}
\usepackage{times}

\usepackage{hyperref}  
\hypersetup{colorlinks=true,urlcolor=blue,citecolor=blue,linkcolor=blue}   
\usepackage[caption=false]{subfig}
\usepackage[shortlabels]{enumitem}
\setlist{nosep} 

\begin{document}

\begin{titlepage}

  
  \title{Context affects student thinking about sources of uncertainty in classical and quantum mechanics}

  \author{Emily M. Stump}
  \affiliation{Laboratory of Atomic and Solid State Physics, Cornell University, 245 East Avenue, Ithaca, NY,
    14853} 
  \author{Matthew Dew}
  \affiliation{Laboratory of Atomic and Solid State Physics, Cornell University, 245 East Avenue, Ithaca, NY,
    14853} 
  \author{Gina Passante}
  \affiliation{Department of Physics, California State University Fullerton, 800 N. State College Blvd., Fullerton, CA,
    92831} 
   \author{N. G. Holmes}
   \affiliation{Laboratory of Atomic and Solid State Physics, Cornell University, 245 East Avenue, Ithaca, NY,
    14853} 


  \begin{abstract}
    Measurement uncertainty is an important topic in the undergraduate laboratory curriculum. Previous research on student thinking about experimental measurement uncertainty has focused primarily on introductory-level students' procedural reasoning about data collection and interpretation. In this paper, we extended this prior work to study upper-level students' thinking about sources of measurement uncertainty across experimental contexts, with a particular focus on classical and quantum mechanics contexts. We developed a survey to probe students' thinking in the generic question ``What comes to mind when you think about measurement uncertainty in [classical/quantum] mechanics?'' as well as in a range of specific experimental scenarios and interpreted student responses through the lens of availability and accessibility of knowledge pieces. We found that limitations of the experimental setup were most accessible to students in classical mechanics while principles of the underlying physics theory were most accessible to students in quantum mechanics, even in a context in which this theory was not relevant. We recommend that future research probe which sources of uncertainty experts believe are relevant in which contexts and how instruction in both classical and quantum contexts can help students draw on appropriate sources of uncertainty in classical and quantum experiments.
  \end{abstract}
  
  \maketitle
\end{titlepage}

\section{Introduction}

Despite ongoing debate about the goals of the undergraduate physics laboratory curriculum, lab instructors consistently articulate the value of learning about experimental measurement uncertainty~\cite{AAPT1997,AAPT2014,Pollard2021}\footnote{Hereafter we will use the word ``uncertainty'' to mean experimental measurement uncertainty unless otherwise specified}. Consequently, student thinking about uncertainty comprises a large portion of research on lab instruction~\cite{HolmesTBD}. Prior work has found that many introductory-level students believe that an experiment should produce a single ``true value'' and that it is unnecessary to consider multiple measurements~\cite{Hull2021,Sere1993,Evangelinos1999,Leach1998,Buffler2001,Allie1998,Lubben2001}. Even when students do acknowledge the need for multiple measurements, they often ignore uncertainty when interpreting their results~\cite{Buffler2001,Leach1998,Pollard2020,Volkwyn2008,Lubben2001,Allie1998,vanKampen2021} or comparing between measurements~\cite{Buffler2001,Lubben2001,Allie1998,Kung2005,Pollard2020,Volkwyn2008,Etkina2008}.
Moreover, students often apply inconsistent reasoning about uncertainty in response to different questions about collecting, analyzing, and interpreting data~\cite{Buffler2001,Allie1998,Lubben2001}.  
Ideas about uncertainty, however, are not limited to undergraduate physics labs. Quantum mechanical systems, for example, have uncertainties that are independent of measurement and these systems involve measurements and uncertainties that are introduced in a distinct way to the measurements and uncertainties of introductory labs. 

While many aspects of student thinking are well-characterized, this study aims to address several key shortcomings of the previous research. In particular, we aim to expand our understanding of student thinking about uncertainty by considering additional student populations, experimental contexts, and aspects of uncertainty, with a particular focus on upper-level students' thinking about sources of uncertainty in classical and quantum experiments. Below we articulate why these approaches to probing student thinking about uncertainty offer critical perspectives to inform instruction.

\subsection{Why sources of uncertainty?}

Identifying sources of uncertainty is integral to modeling experiments. For example, in the Modeling Framework for Experimental Physics~\cite{DounasFrazer2018,Zwickl2015}, the modeler must consider limitations, simplifications, and assumptions as well as principles and concepts when constructing models of both the physical and measurement systems of an experiment. In doing so, the modeler must interrogate sources of uncertainty in their measurements. Considering these sources directly informs experimental procedures and improvements, such as what equipment to use and how much data to collect. For example, if a student attributes uncertainty to the precision limits of a measuring device they may choose to use a more precise device. On the other hand, if students attribute uncertainty to ``human error'' that can and should be eliminated~\cite{Holmes2015,Hu2018,Sere2001,Hull2021}, they may discount the importance of taking multiple measurements or they may ignore uncertainty when interpreting or comparing experimental results~\cite{Pillay2008,Evangelinos2002}.

Previous work has connected student thinking about sources of uncertainty to procedural reasoning~\cite{Etkina2008,Sere1993,Kung2005,Holmes2015}. For example, S\'{e}r\'{e} \textit{et al.}~\cite{Sere1993} observed that students could identify multiple sources of uncertainty when conducting an optics experiment but tended to only consider a single source, such as the precision of the ruler, in quantifying the uncertainty in their own measurements. Similarly, Kung~\cite{Kung2005} argued for the importance of distinguishing external and internal variation in a lab curriculum to help students productively compare data sets. Therefore, assessments asking students about sources of uncertainty provides an additional lens on student thinking about uncertainty beyond but complementary to procedural reasoning.  

\subsection{Why different experimental contexts?}
\label{sec:variables}

Previous research on student thinking about uncertainty has largely focused on introductory-level students and simple classical experiments with a single-value outcome predicted by physics theory. Research has shown, however, that student thinking about experimental physics may vary greatly depending on the context in which it is probed~\cite{Leach2000}. Thus, students may think differently about uncertainty outside the context of a simple, deterministic classical experiment.

One approach to understanding this context-dependence is through the lenses of concept images~\cite{Tall1981,Schermerhorn2021} and cognitive accessibility~\cite{Heckler2018}. The various pieces of knowledge~\cite{Hammer2000} that students draw on in different contexts collectively make up a student's concept image~\cite{Tall1981,Schermerhorn2021} of measurement. Although students may have many pieces of knowledge related to measurement in their concept images, they may or may not draw on these ideas in a given context. We refer to knowledge pieces present in students' concept images as being ``available,'' while we refer to knowledge pieces activated by a given context as being ``accessible'' in that moment~\cite{Heckler2018}.

\subsubsection{Adding quantum mechanics contexts}

Quantum mechanics is a natural alternative context to study, as uncertainty in this context can refer either to aspects of the experiment or to uncertainty that is inherent in the physics theory, for example, the Heisenberg uncertainty principle. This duality complicates students' reasoning about uncertainty in quantum experiments~\cite{Ayene2011,KrijtenburgLewerissa2017,Singh2015,Zhu2012a,Bouchee2022,Johnston1998}. For example, Ayene \textit{et al.} found that undergraduate physics students tended to describe the Heisenberg uncertainty principle in terms of mistakes made by an experimenter, rather than as an inherent aspect of the phenomenon of interest~\cite{Ayene2011}. Thus, students may not have a robust understanding of uncertainty inherent to the theory of quantum mechanics and how it differs from experimental considerations that may be present in both classical and quantum experiments. 



Furthermore, students may be more likely to see classical experiments as deterministic and able to produce a single ``true value'' than a quantum experiment. For example, Dreyfus \textit{et al.} asked students the extent to which they agreed with the statement ``It is possible for physicists to carefully perform the same measurement in a [classical/quantum] physics experiment and get two very different results that are both correct''~\cite{Dreyfus2019}. The researchers found that students were more likely to agree with the statement in the quantum context than the classical context, implying that variation in results is more acceptable and even expected within a quantum physics experiment. 

In our previous work, we expanded these earlier findings to probe student thinking about uncertainty within specific classical and quantum experiments through a series of interviews with upper-level students~\cite{Stein2019,Stump2020}. In these interviews, students were presented with classical and quantum experimental scenarios and corresponding fictitious data distributions and were asked to discuss the uncertainty present in the results. This work found that students primarily attributed uncertainty in a classical projectile motion experiment to limitations in the experimental apparatus. In contrast, they attributed uncertainty in a quantum single-particle single-slit experiment to inherent aspects of the physics theory. These results were consistent when students were asked a more generic question ``What comes to mind when you think about measurement uncertainty in [classical/quantum] mechanics?'' In the current study, we aim to further probe which aspects of students' concept images related to uncertainty are accessible based on the physics paradigm (classical or quantum) of an experiment with a larger sample of students and a greater range of experimental scenarios.

\begin{figure*}[t]
  \centering
  \includegraphics[width=\textwidth]{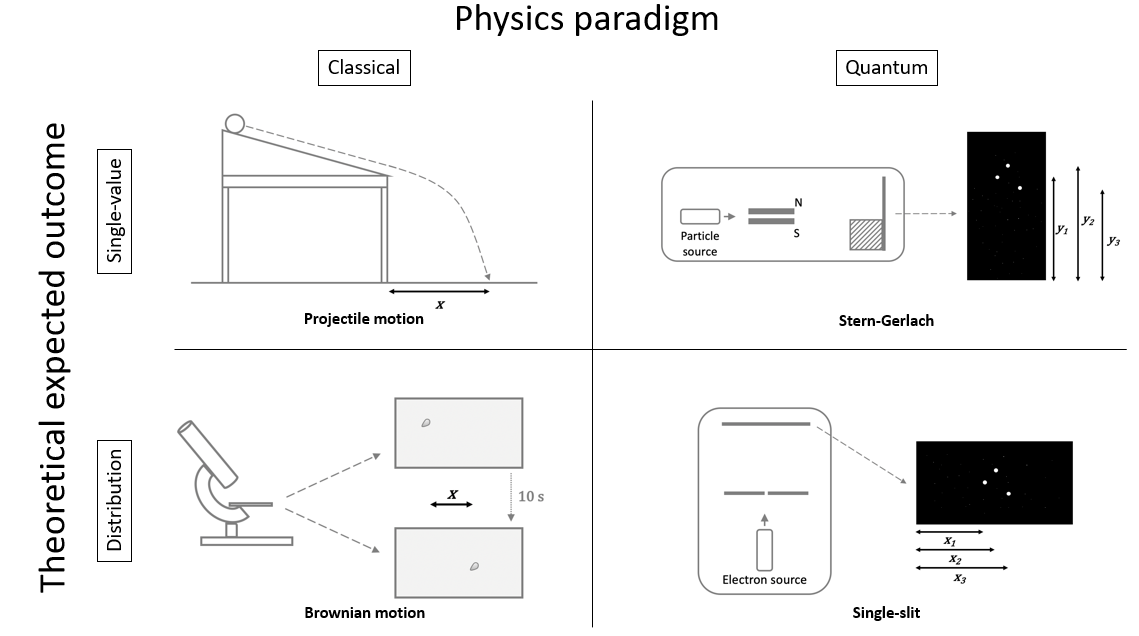}
  \caption{The four experimental scenarios included in the survey: projectile motion, Stern-Gerlach, Brownian motion, and single-slit. These four experiments span the possible combinations of physics paradigm (classical or quantum) and theoretical expected outcome (single value or distribution).}
  \label{fig:exp_grid}
\end{figure*}

\subsubsection{Considering additional factors beyond classical/quantum}

The Dreyfus \textit{et al.} study~\cite{Dreyfus2019} raises another key issue in understanding the role of experiment context in student thinking about uncertainty: classical experiments are often portrayed as deterministic while quantum experiments are portrayed as probabilistic. How then, does the expected outcome of an experiment, namely whether the physics theory predicts a single-value outcome (e.g. the projectile motion experiment) or a distribution of outcomes (e.g. the single-slit experiment) affect which pieces of knowledge are accessible to students? In our previous work, we conflated these two variables (expected outcome and physics paradigm) by reporting on students' responses to a single-value classical experiment and a distribution quantum experiment~\cite{Stein2019, Stump2020}.

Other work has suggested that students may think similarly about experiments with an expected distribution of outcomes regardless of whether the underlying phenomenon is due to classical or quantum physics. Baily and Finkelstein asked modern physics students to compare the double-slit experiment with a classical Plinko game, in which a marble drops through a series of pegs, landing in a random bin at the bottom~\cite{Baily2009a}. They found that most students viewed the two systems as similar in that both would produce a distribution of possible position measurements. Moreover, many of the students were unable to articulate how the two systems might be different based on the presence or absence of quantum-mechanical effects. In the current study, we aim to disentangle the effect of physics paradigm from the effect of theoretical expected outcome of an experiment.

Finally, students may also respond differently depending on whether a question is generic or focuses on a specific experimental scenario. For example, Leach \textit{et al.}~\cite{Leach2000} observed that introductory-level students drew on different lines of reasoning when interpreting data from a superconductivity experiment and when answering generic questions outside of an experimental scenario. In our previous interview study, however, we found that student responses were similar between specific experimental scenarios and the generic question ``What comes to mind when you think about measurement uncertainty in [classical/quantum] mechanics?'' Thus, we aim here to further probe the relationship between student responses to specific experimental scenarios and to generic questions to better identify when context matters for knowledge accessibility.

\subsection{Research aims}

Overall, in this work we aim to characterize what sources of uncertainty upper-level students from a variety of institutions identify across a range of variables: (1) physics paradigms (namely, classical and quantum contexts), (2) theoretical expected outcomes (namely, single values and continuous distributions), and (3) generic and experiment-specific questions. We draw on the Modeling Framework for Experimental Physics~\cite{DounasFrazer2018,Zwickl2015} to characterize the sources as either principles or limitations and evaluate students' responses through a lens of the availability and accessibility of ideas~\cite{Heckler2018}. This work is not meant to evaluate a specific instructional lab pedagogy but rather to characterize diverse students' thinking about uncertainty as a foundational concept that is relevant across a range of physics courses.

\section{Methods}


\subsection{Survey development}
\label{sec:survey}

We created a survey to probe student thinking about uncertainty in a variety of contexts. The structure of the survey was based on the interview protocol from Stein \textit{et al.}~\cite{Stein2019}. Students were presented with several experimental scenarios and asked to answer a series of questions about uncertainty for each scenario. Afterward, students were asked ``generic'' questions about measurement uncertainty in classical and quantum mechanics that were not tied to a particular experimental context. 

\subsubsection{Experimental scenario items}
We aimed to probe how student thinking about sources of uncertainty varied based on both physics paradigm (classical or quantum) and theoretical expected outcome (single value or distribution). Thus, we developed four experimental scenarios that cover all possible combinations of paradigm and theoretical outcome (see Fig.~\ref{fig:exp_grid}): a projectile motion experiment in which a ball rolls down a ramp (classical, single value), a Stern-Gerlach experiment with one particle stream blocked (quantum, single value), a Brownian motion experiment (classical, distribution), and a single-particle single-slit experiment (quantum, distribution). The projectile motion setup was based on the experimental scenario from the Physics Measurement Questionnaire~\cite{Allie1998}. The other three scenarios, which focus on upper-level physics content, were chosen because they (or similar experiments) commonly appear in theory-oriented physics textbooks and courses. 
By choosing experiments that commonly appear in theory courses, we hoped to provide students with experimental scenarios they would have seen before, although not necessarily in a laboratory context. At the end of the survey, we also asked students a Likert question about their comfort with each experiment they saw in the survey: 
\begin{quote}
``How comfortable were you with the [projectile motion/Stern-Gerlach/Brownian motion/single-slit] experiment presented in this survey?''
\end{quote}
with answer options ``Very comfortable,'' ``Somewhat comfortable,'' ``Somewhat uncomfortable,'' and ``Very uncomfortable.''

\begin{figure*}[t]
\centering
  \fbox{\includegraphics[width=.75\textwidth]{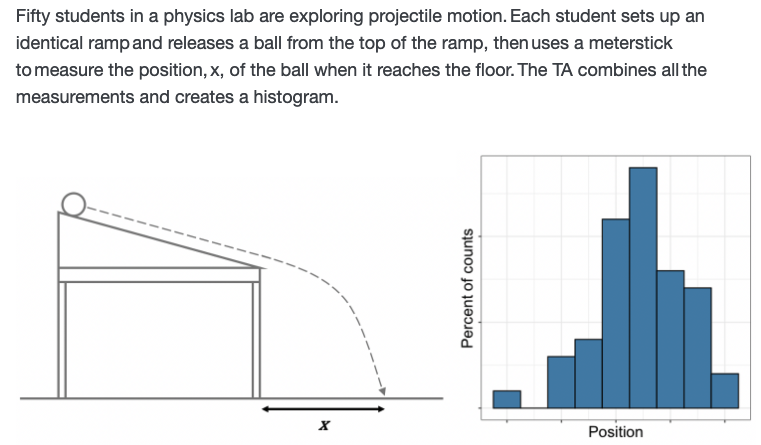}}
  \caption{The descriptive text, diagram, and histogram of fictitious experimental data for the projectile motion scenario.}
\label{fig:PM}
\end{figure*}

Each experimental context included a text description, a diagram of the experiment (see Fig.~\ref{fig:exp_grid}), and a histogram of fictitious experimental results. As an example, the projectile motion context is shown in Fig.~\ref{fig:PM}
(the full descriptions for the other experiments may be found in Fig.~\ref{fig:scenarios_full} in the appendix). Through the descriptions and diagrams, we aimed to minimize the impact experimental setup complexity may have on student responses. Each description was intended to be both vague enough so as not to limit student thinking while also specific enough to ensure that students could understand how the experiment might be conducted. Where possible we attempted to ensure that the descriptions were similar so as not to bias student responses based on differences in experimental methods. For example, each scenario involves a class of fifty students measuring a distance using a meterstick, the students are always measuring individual particles, and the measurements are displayed as identical histograms (as shown in Fig.~\ref{fig:PM}). The graphs also do not display specific units or numeric values so as not to limit the sources of uncertainty students might consider based on the absolute spread in measurement values.

For each experiment, participants were asked to generate a list of sources of uncertainty in response to the prompt:
\begin{quote}
 ``What is causing the shape of the distribution? List as many causes as you can think of.''
 \end{quote}
The survey was structured so that participants could list each source of uncertainty in a separate text box; after listing each source, students could generate an additional text box, up to a maximum of ten. The survey continued with several additional questions about each context, but these are not analyzed in this work.

\subsubsection{Generic items}

After participants had considered the individual experiments, they were then presented with two open-response generic questions probing their thoughts about  uncertainty in general, first in classical mechanics and then in quantum mechanics: 
\begin{quote}
``What comes to mind when you think about measurement uncertainty in [classical/quantum] mechanics?''
\end{quote}
This question was lifted verbatim from the prior interview protocol~\cite{Stein2019} and students had an open text box to respond to each question.

\subsection{Data collection}
\label{sec:data_collection}

We piloted the full survey in interviews with twelve students from Cornell University and California State University Fullerton to check that the questions were unambiguous and that students were interpreting them as we had intended. These pilot interviews led us to decrease the number of questions asked for each experimental scenario and to modify the wording of several questions to improve clarity. We also decreased the number of experimental scenarios each participant saw in the survey from four to two to reduce time required to complete the survey. Each participant responded first to the projectile motion experiment and then to one of the other three scenarios, chosen at random. This choice allowed us to gather information about upper-level student thinking about a simple intro-level experiment while ensuring that all respondents had as similar an experience as possible filling out the survey. Because this last change was implemented after we had begun administering the survey, our results include a small number of participants who responded to all three upper-level experimental scenarios.


The variability in which upper-level experiment each student saw may have impacted their responses to the generic questions about uncertainty in classical and quantum mechanics. For example, students who saw a classical experiment (Brownian motion) may have responded differently to the generic questions than students who saw a quantum experiment (Stern-Gerlach or single-slit).

We collected survey responses during the second half of the Fall 2020 semester and the second half of the Spring 2021 semester. We recruited students from five different institutions, including private universities, public universities, primarily white institutions, and a Hispanic-serving institution. For more details about the participants' demographic information, see Table~\ref{ta:demographics} in the appendix. At all institutions, we recruited students who had previously taken or were currently taking a quantum mechanics course. Most of these students (84\%) indicated that they were physics majors; the others were either engineering, math, or other physical sciences majors. Students were recruited through faculty at their individual institutions. The survey was administered online through Qualtrics. Students who participated were entered in a draw for a \$25 gift card or were given course credit by their instructors.

In total, we received completed survey responses from 150 students who consented to take part in this study. The number of sources listed by each student for each experimental scenario varied from one to nine sources (Table~\ref{ta:source_numbers}).
\begin{table}[tb] 
  \caption{Descriptive statistics of the sources listed by each respondent for each experimental scenario show little difference among the scenarios. The median number of sources are reported per student and the middle 50\% of number of sources provides the bounds of the interquartile range.}
  \label{ta:source_numbers}
  \begin{ruledtabular}
    \begin{tabular}{p{.13\textwidth}p{.11\textwidth}p{.08\textwidth}p{.1\textwidth}}
    \textbf{Scenario} & \textbf{Respondents} & \textbf{Median number of sources} & \textbf{Middle 50\% of number of sources} \\
    \hline
    Projectile motion & 150 & 3 & [2, 4] \\
    Stern-Gerlach & 59 & 2 & [2, 3] \\
    Brownian motion & 59 & 3 & [2, 3]\\
    Single-slit & 60 & 2 & [1, 3]\\
    \end{tabular}
  \end{ruledtabular}
\end{table}

\begin{figure}
  \centering
  \includegraphics[width=.49\textwidth]{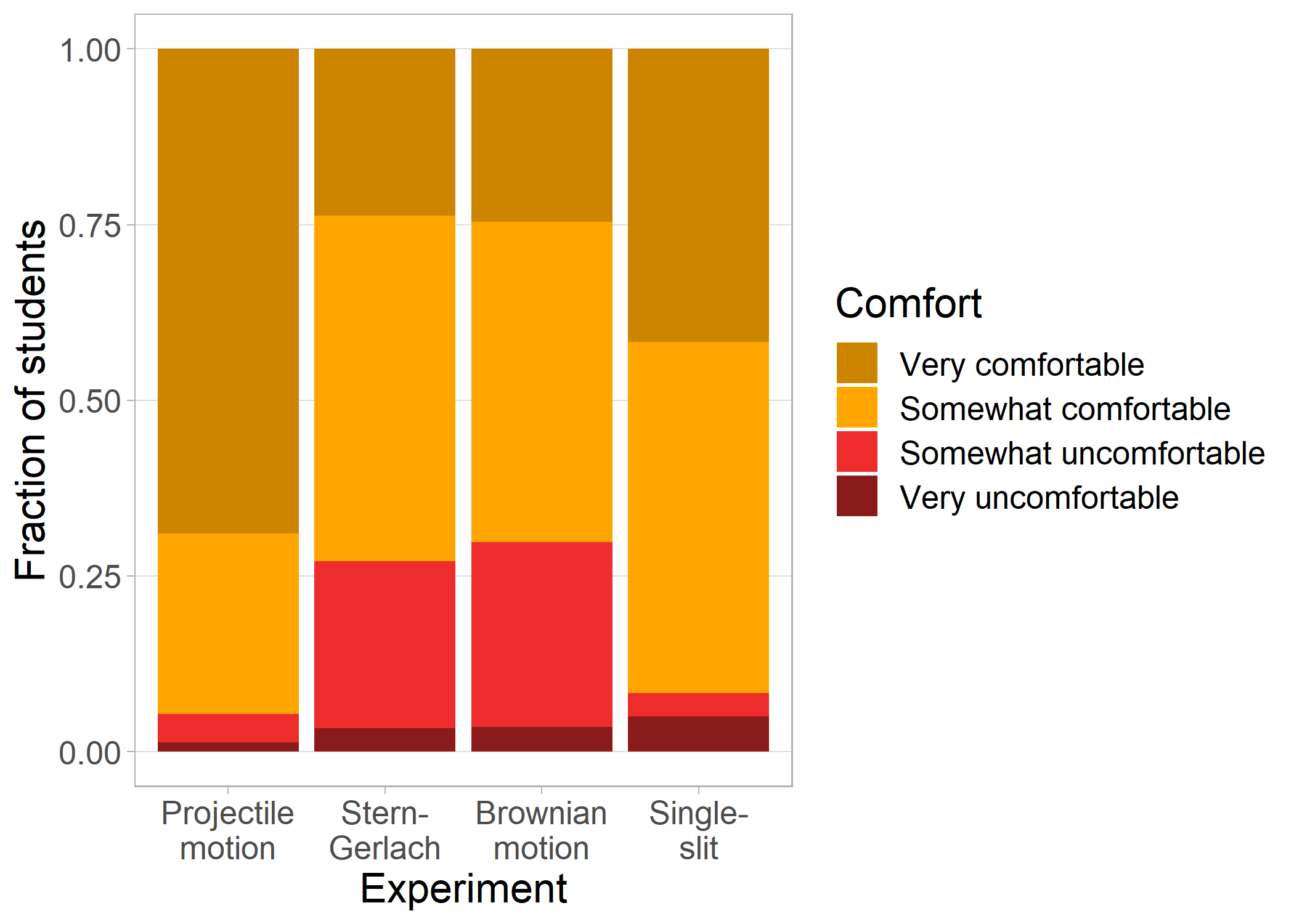}
  \caption{Respondent comfort with the experimental scenarios in the survey.}
  \label{fig:comfort}
\end{figure}

Almost all students reported being comfortable with the projectile motion and single-slit experiments (see Fig.~\ref{fig:comfort}).
Approximately 25\% of student respondents who saw the Stern-Gerlach and Brownian motion experiments, however, reported being uncomfortable with these scenarios. Although these students were likely less familiar with the experiments from their previous course experience, there was not a statistically significant difference in the number of sources they listed relative to their comfortable counterparts. 

\begin{table*}[tb] 
  \caption{Definitions and examples of the limitations and principles codes.}
  \label{ta:coding_scheme}
  \begin{ruledtabular}
    \begin{tabular}{p{0.08\textwidth}p{0.4\textwidth}p{0.45\textwidth}}
    \multicolumn{1}{c}{\textbf{Code}} & \multicolumn{1}{c}{\textbf{Definition}}& \multicolumn{1}{c}{\textbf{Examples}}\\
      \hline
      &&\\
    Limitations & Variability owing to our inability to perfectly model and measure a physical system in a real-world experiment & \textit{``Imprecision due to human error reading the place where the ball hits the floor''} (Projectile motion)\\
    && \textit{``Misalignment of the particle beam direction with the magnet and detector''} (Stern-Gerlach)\\
    &&\textit{``Differences in particle characteristics (size, energy, etc)''} (Brownian motion)\\
    &&\textit{``Different slit sizes''} (Single-slit)\\
    &&\\
    Principles & Variability inherent in the theoretical abstraction of an experiment or statistical modeling of measurement & \textit{``Repetition of experiment gives the gaussian like curve''} (Projectile motion) \\
    && \textit{``Quantum uncertainty''} (Stern-Gerlach) \\
    && \textit{ ``Brownian motion is a random phenomena of moving particles. Thus how much a particle moves is randomizes and the distribution should be a normal gaussian which captures this random process.''} (Brownian motion)\\
    & & \textit{ ``Randomness from quantum mechanical probability distribution''} (Single-slit)\\
    \end{tabular}
  \end{ruledtabular}
\end{table*}

\subsection{Coding scheme}
\label{sec:coding_scheme}

We developed a coding scheme to classify student-listed sources of uncertainty across the experimental scenarios and their responses to the generic questions. Although responses to the generic questions were not necessarily exclusively focused on sources of uncertainty, applying the same coding scheme to both types of questions allowed us to compare responses across the physics paradigms generically and when embedded in experimental scenarios.


The coding scheme built on our previous work~\cite{Stein2019,Stump2020} and drew on the Modeling Framework for Experimental Physics~\cite{Zwickl2015,DounasFrazer2018}. Within the Modeling Framework, experimenters consider separately the \emph{principles} underlying the physical system of interest and the experimental measurement system as well as the \emph{limitations} in their modeling of the physical and measurement systems. We developed these two concepts into two codes (limitations and principles) and used student responses to our survey to further refine our definitions as exemplified in the data. The code definitions and examples of each code from different experiments are described in Table~\ref{ta:coding_scheme}.

Sources coded as limitations focus on imperfections in a specific experimental setup, owing to the fact that that we cannot perfectly model and measure a physical system in a real-world experiment. The limitations encompass a broad range of sources of uncertainty, ranging from students being careless in how they set up an experiment to the limits on the precision of a measuring device. They include limitations in both the measurement system (e.g. human error reading the ruler) and the physical system (e.g. air resistance).

The principles code includes all sources of uncertainty that are based in inherent variability that is part of the theoretical physics abstraction of an experiment or is a statistical principle of experimental measurement. The theoretical physics aspect of this code includes phenomena such as electron diffraction in the single-slit experiment or random particle motion in the Brownian motion experiment that inherently cause variability in the system (see Stern-Gerlach, Brownian motion, and single-slit examples in Table~\ref{ta:coding_scheme}). It also includes general physics principles such as the Heisenberg uncertainty principle or chaos. 

The statistical modeling aspect of the principles code includes the idea that the experimental measurement process is inherently random and should be modeled using statistical probability distributions, regardless of the specifics of the physical or measurement systems, for example, \textit{``Multiple random measurements result in a gaussian distribution''} (see projectile motion example in Table~\ref{ta:coding_scheme}). Previous work has advocated teaching students this view of uncertainty as an inherent principle of measurement modeled with probability distributions because it helps students become more expertlike in their interpretation of experimental results~\cite{Evangelinos1999,Evangelinos2002,Allie2003,Pillay2008,Buffler2008}, so we included this concept in our definition of the principles code. In practice, only 9\% of principles codes focused on this principle of measurement, so the principles code as represented in our data is almost exclusively concerned with the theoretical physics abstraction of an experiment.

Overall, our coding scheme captured the majority of student responses, with limitations and/or principles codes being applied to 93\% of the sources listed for the specific experiments and 75\% of the responses to the generic uncertainty questions. Most of the remaining responses that mentioned sources of uncertainty were vague, such as \textit{``experimental error''} or \textit{``random errors,''} which we could not distinguish as either reflecting principles or limitations. 

Two coders independently coded a random sample of 20 student-listed sources across the experimental scenarios and refined and clarified the code definitions by discussing and resolving disagreements. The coders then independently coded a separate random sample of 40 student-listed sources. Inter-rater reliability was quantified using Cohen's kappa, a measure of how well two coders agree beyond the level expected due to random chance~\cite{Cohen1960}. The kappa value for this second sample was 0.85, which indicates almost perfect agreement~\cite{Landis1977}. 

\subsection{Data analysis}

Student responses for both the experimental scenarios and the generic questions were analyzed based on whether a student received at least one code of limitations and, separately, on whether a student received at least one code of principles. We examined the existence of each code, rather than how many times each student received that code, for two reasons. First, we did not expect that students would list a similar number of limitations and principles for each experimental scenario. In most cases, it would be appropriate to list only a single theoretical physics and/or statistical modeling principle for each scenario, but there would be many possible limitations students might have listed. Thus, comparing the number of limitations to the number of principles listed would tell us more about the nature of the codes than it would about student thinking. In addition, we intended to directly compare student responses to the generic questions and to the experimental scenarios. However, the structure of student responses differed between these two question types, with students writing out a single response to the generic question but listing multiple sources of uncertainty in separate text boxes for the experimental scenarios. By looking only for the presence of each code, rather than how many times it occurred within a single student's response, we eliminate any bias that might emerge based on how we chose to segment the generic question responses.


While our data could be analyzed with, for example, $\chi^2$ tests of distinguishability, the analysis involves a large number of comparisons and so we risk finding apparent statistical significance due only to chance. Given that relying on $p$-values can be problematic~\cite[see, for example, ][]{cohen_earth_1994, cumming_new_2013, nosek_preregistration_2018}, we choose to only make qualitative interpretations of the proportions throughout the manuscript. Overall, we are looking for large-scale distinctions in the data that are apparent with the provided error bars and do not comment on the possible distinguishability of small effects.

\subsection{Limitations}

Our study design presents multiple limitations, some of which we sought to mitigate and some of which motivate future work. First, any survey study risks multiple forms of sampling bias. For example, students with higher course grades are more likely to respond to optional course surveys than students with lower course grades~\cite{Nissen2019} and students with significant demands on their time may not respond to a survey or may stop responding to a long survey partway through. We sought to mitigate against these forms of sampling bias by providing moderate incentives for participation and making the survey as short as possible (such as giving each student two of the four scenarios, rather than all four). Another form of sampling bias relates to the demographic characteristics of the surveyed students, which often do not reflect the diversity of physics students nationally~\cite{Kanim2020}. While we intentionally sought to obtain participants from a range of institutions and academic levels, the majority of our participants identified as men, were enrolled at PhD-granting institutions, and identified as White or Asian/Asian American. Future work should seek a more diverse sample of students, such as explicitly recruiting students at minority-serving institutions. The results below should be interpreted with these limitations to generalizability in mind.

Another limitation exists within our coding scheme. Our analysis broadly examines the overarching ideas of principles and limitations in student thinking about uncertainty. Both of these constructs, however, are multifaceted and contain many subcategories and ideas. In the coding scheme, for example, we clarified the role of principles in terms of the theoretical abstraction of the system and the statistical modeling of measurement. Our analysis, however, does not distinguish these two forms of principles in student thinking. In the current study, ideas about statistical modeling of measurement were quite rare, so our interpretations of the principles code largely represent the role of the theoretical abstraction of the system. The overarching ideas of principles and limitations were appropriate to answer our research question about characterizing the range of student thinking across a range of contexts. Future work, therefore, should disaggregate this code to examine further nuances in student thinking across contexts, especially when the goal is to evaluate the role of instruction on student thinking. We also note that the limitations code is incredibly broadly defined and would benefit in subsequent work from more nuanced examinations of how students evaluate limitations. Such nuance is beyond the scope of this analysis. 

Lastly, we acknowledge limitations in our decision to use only qualitative interpretations of the proportions, rather than statistical comparisons. While this decision is to mitigate against Type I errors (seeing effects due to chance when making multiple comparisons), our analysis likely misses small effects in the data that may be meaningful for other research questions. Future work with more narrow research questions may wish to use statistical comparisons, particularly if fewer comparisons are being made.

\section{Results}

In this section, we report on the sources of uncertainty students listed and how these varied by context. We start by discussing variation in student responses across experimental contexts and then presenting differences in student responses to the generic uncertainty questions.

\subsection{Similarities and differences across experiments}
\label{sec:RQ2_exp}

\begin{figure}
  \includegraphics[width=.49\textwidth]{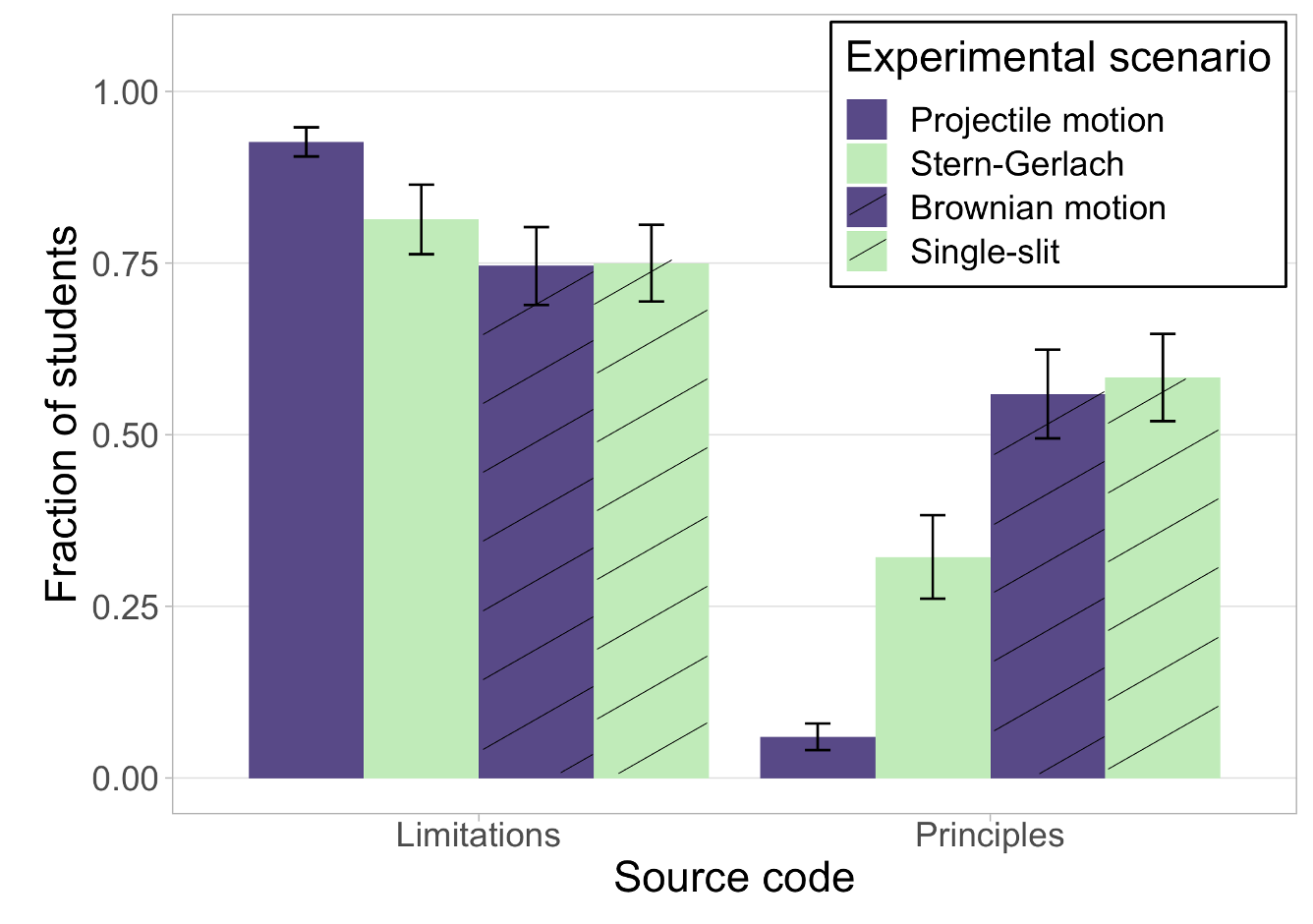}
  \caption{Codes applied to sources of uncertainty listed for each experiment in response to the prompt ``What is causing the shape of the distribution? List as many causes as you can think of.'' Physics paradigm is denoted by color (purple = classical, green = quantum), while theoretical expected outcome is denoted by pattern (solid = single value, striped = distribution). Error bars represent the standard error on the proportions.}
    \label{fig:top_experiment}
\end{figure}

\subsubsection{Overall response patterns}

For each of the experiment contexts, students were asked to list as many sources of uncertainty (``causes of the shape of the distribution'') as they could think of. 
For all four experiments, most students (ranging from 75\% to 93\%) listed at least one source coded as limitations (see Fig.~\ref{fig:top_experiment}). Some of these limitations appeared across all four experiments, such as ruler precision or human error. Others were experiment-specific, such as \textit{``deviation in the initial release of the ball''} for projectile motion or \textit{``fluctuations in the surrounding EM field between trials''} in the single-slit experiment. For example, for the Stern-Gerlach experiment, students listed limitations sources such as \textit{``position of the magnet,''} \textit{``speed of emitted particles,''} and \textit{``particle source doesn't send out particles directly horizontally.''} For the Brownian motion experiment, limitations sources included \textit{``temperature tiny differences,''} \textit{``any perturbation in the placement of the [microscope] slide,''} and \textit{``tapping the microscope or other jitters from the outside world.''}

Students listed sources coded as principles less frequently. For the three upper-level experiments, between 32\% and 58\% of students listed at least one source coded as principles (see Fig.~\ref{fig:top_experiment}). These tended to be specific to the experiment's physics paradigm. For Brownian motion, principles tended to focus on the inherently random nature of Brownian motion, such as \textit{``random walks distribution''} or simply \textit{``the randomness of Brownian motion.''} For the Stern-Gerlach and single-slit experiments, the principles tended to focus on quantum-mechanical principles, such as the Heisenberg uncertainty principle, or other more vague descriptions of quantum-mechanical uncertainty, such as \textit{``quantum randomness''} and \textit{``some quantum probability phenomenon.''} Students rarely mentioned principles as sources of uncertainty in the projectile motion experiment (6\% of students). When students did mention principles for projectile motion, they primarily alluded to the inherent statistical nature of experimental measurement, for example \textit{``Multiple random measurements result in a gaussian distribution.''}

\subsubsection{Responses based on theoretical expected outcome and physics paradigm}

Although the overall trend that more students list limitations than principles is consistent across all four experiments, we observed some variation in the fraction of students who listed principles sources for each experiment. As discussed in Sec.~\ref{sec:variables}, we were interested in how student responses varied based on the theoretical expected outcome of the experiment (single value or distribution) and on the physics paradigm of the experiment (classical or quantum). For each comparison, we simultaneously control for the other variable to disentangle the two potential effects (e.g., to compare based on theoretical expected outcome, we compare between the two classical experiments and then between the two quantum experiments).

When comparing student responses based on the theoretical expected outcome of an experimental scenario, we found that more students listed a principles source for the distribution experiments (striped) than they did for the single-value experiments (solid; see Fig.~\ref{fig:top_experiment}).  For the classical paradigms, 56\% of students listed a principles source for Brownian motion (distribution) compared with 6\% for projectile motion (single value). For the quantum paradigms, 58\% of students listed a principles source for single-slit (distribution) compared with 32\% for Stern-Gerlach (single value). 

When comparing student responses based on the physics paradigm of an experimental scenario, however, we did not observe consistent differences between the classical (purple) and quantum (green) experimental scenarios. For the single-value experiments, the Stern-Gerlach experiment (quantum) had 32\% of students listing a principles source compared with 6\% of students for the projectile motion experiment (classical). The two distribution experiments, on the other hand, had nearly the same fraction of principles sources (56\% for Brownian motion (classical) and 58\% for single-slit (quantum); see Fig.~\ref{fig:top_experiment}). Thus, students do not consistently draw more heavily on principles in a quantum context than in a classical context.

\subsubsection{Responses based on student comfort}

One possible confounding variable in comparing student responses across experiments is students' familiarity with the experiments. Nearly all respondents indicated that they were comfortable with the projectile motion and single-slit experiments. In contrast, around 25\% of students who responded to the Stern-Gerlach and Brownian motion experiments indicated that they were uncomfortable with these experiments (see Fig.~\ref{fig:comfort}). We therefore explored how student responses varied based on their reported comfort with these two experiments (see Fig.~\ref{fig:BMSG_comfort} in the appendix). Although we observe small differences in the fraction of comfortable and uncomfortable students who listed principles sources for each scenario, these differences are too small relative to their uncertainty to draw any conclusions.

\subsection{Generic questions by physics paradigm}

\begin{figure}
  \centering
  \includegraphics[width=.49\textwidth]{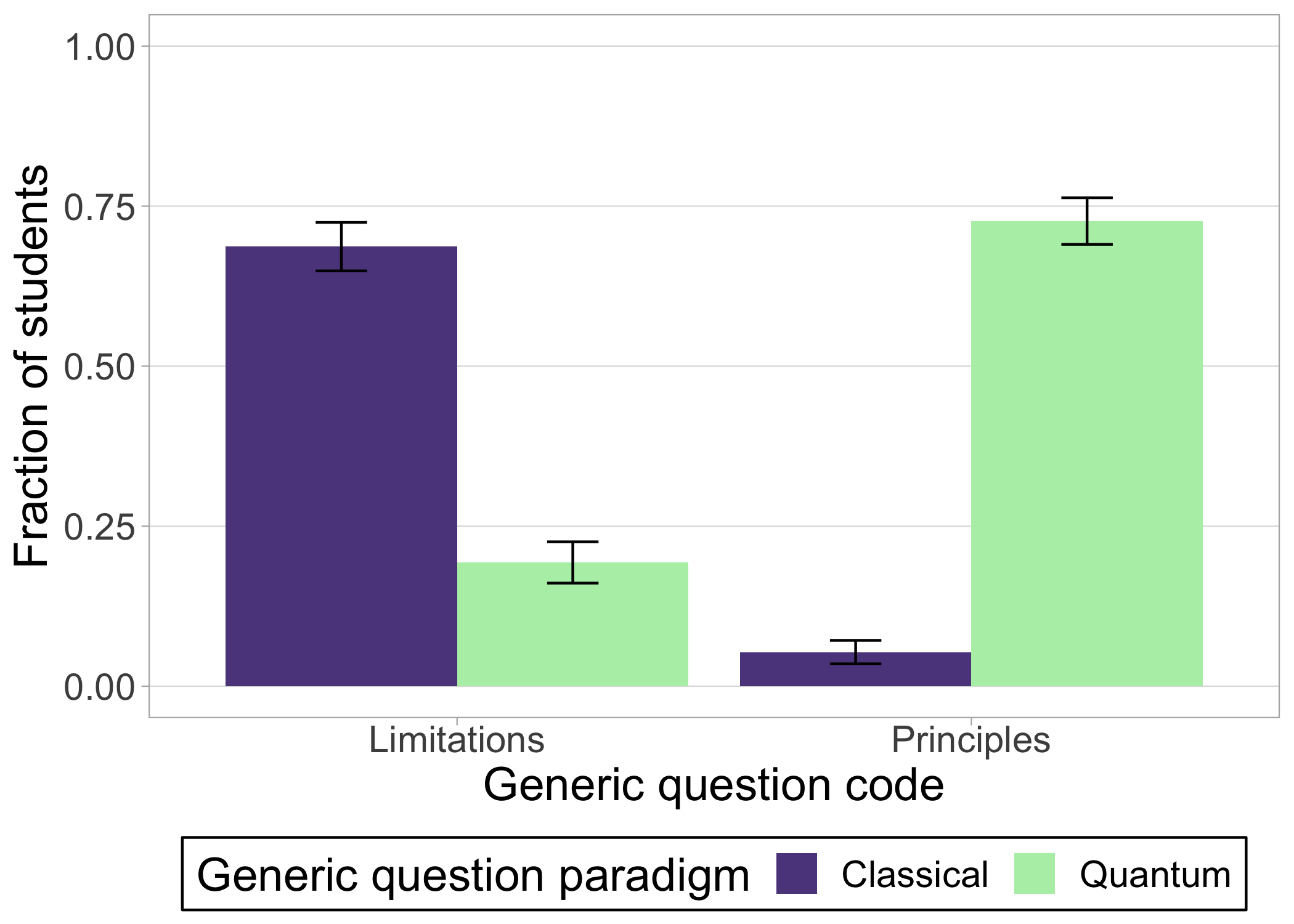}
  \caption{Codes applied to student responses to the question ``What comes to mind when you think about measurement uncertainty in [classical/quantum] mechanics?'' Error bars represent the standard error on the proportions. Physics paradigm is denoted by color (purple = classical, green = quantum).}
  \label{fig:top_gen}
\end{figure}

After considering the specific experiments, students were asked to describe what they think about when considering uncertainty first in classical mechanics and then in quantum mechanics. Overall, in response to the generic questions, students focused primarily on limitations sources of uncertainty in a classical context and principles in a quantum context. 

In response to the classical question, 69\% of students mentioned at least one limitations source of uncertainty, such as \textit{``human errors,''}, \textit{``friction,''} and \textit{``measuring tools.''} For example, one student specifically described examples of limitations sources of uncertainty: \textit{``Classical errors, such as a bent straight edge or friction. The error of a timer or a measurement not being instantaneous.''} Another more tangentially touched on the idea of instrumental uncertainty: \textit{``I think of those annoying rules they taught us in high school chemistry/physics where you had to start measuring from the 1st inch of the ruler and the measurement uncertainty is twice the size of the smallest notch of your ruler/beaker or something like that.''} 

In contrast, in response to the quantum question, only 19\% of students mentioned limitations. Although more students listed limitations for the classical question than for the quantum question, the nature of those limitations sources was similar in both paradigms. The minority of students who listed limitations for the quantum question tended to repeat the sources they had listed for the classical question. For example, one student wrote, \textit{``Classical measurement uncertainty plays a role in QM experiments, too.''}

For the principles code, the opposite was true: most students mentioned at least one principles source of uncertainty in response to the quantum question (73\%), with most of these students mentioning the Heisenberg uncertainty principle or other properties of quantum mechanics, such as \textit{``the wave like nature of particles.''} For example, one student wrote \textit{``The inherent probabilistic nature of quantum mechanics.''} Another commented that uncertainty is \textit{``more or less supposed to be there since QM is a statistical field at the scale that we care about.''} Very few students, however, mentioned principles in response to the classical question (5\%).

\subsection{Interaction between generic questions and experiments}
\label{sec:priming}

In the survey, students responded first to the projectile motion scenario and then to a randomly chosen upper-level scenario before responding to the generic uncertainty questions. We wanted to check, therefore, for any trends between which experimental context each student saw (particularly regarding physics paradigm) and how they answered the generic questions. We observed no distinguishable differences in students' listing of limitations and principles based on what upper-level experiment students saw (see Fig.~\ref{fig:gen_exp} in the appendix).

\section{Discussion}

In this study, we probed student thinking about sources of uncertainty in classical mechanics and quantum mechanics. We observed a clear split in student thinking based on physics paradigm when considering uncertainty in classical and quantum generically, with students primarily mentioning limitations in the experimental setup in classical mechanics and principles of theoretical physics in quantum mechanics (Fig.~\ref{fig:top_gen}). When students considered specific experimental scenarios, however, this same split did not emerge (Fig.~\ref{fig:top_experiment}). Regardless of physics paradigm, most students listed at least one limitations source and students' likelihood of listing a principles source was not explained by physics paradigm alone.

We interpret these results through the lens of cognitive accessibility~\cite{Heckler2018} to understand why students responded so differently in different contexts. Previous research has indicated that students can express very different ideas about experimental measurement in different contexts~\cite{Leach2000}. The various pieces of knowledge~\cite{Hammer2000} that students draw on in different contexts collectively make up a student's concept image~\cite{Tall1981,Schermerhorn2021} of measurement. Although students may have many pieces of knowledge related to measurement in their concept images, they may or may not draw on these ideas in a given context. We refer to knowledge pieces present in students' concept images as being ``available,'' while we refer to knowledge pieces activated by a given context as being ``accessible'' in that moment~\cite{Heckler2018}. Within the context of our survey, we can conclude that any piece of knowledge students mention in a response is both accessible and available to that student. If students do not address a particular knowledge piece, however, we cannot know whether that piece of knowledge is available but not accessible or whether that piece of knowledge is simply unavailable.

We consider each student's knowledge about uncertainty in classical mechanics and uncertainty in quantum mechanics to be separate concept images (though with the potential for significant overlap between them). Based on student responses to the survey, we discuss whether limitations and principles are available and/or accessible to students within quantum and classical mechanics. We first describe the types of sources that are primarily accessible in each paradigm: principles in quantum mechanics and limitations in classical mechanics. We then discuss the less accessible and/or available types of sources within each paradigm (limitations in quantum mechanics and principles in classical mechanics).

\subsection{Primary accessible sources of uncertainty: limitations in classical and principles in quantum}

We observed that limitations (e.g. human error, instrumental precision, friction) are the most consistently accessible sources of uncertainty in classical mechanics, both generically (69\% of students mentioned at least one limitations source) and in the context of specific experimental scenarios (93\% mentioned limitations for projectile motion and 75\% for Brownian motion). 
Thus, we infer that across contexts, limitations sources of uncertainty are consistently accessible to students in classical mechanics. The accessibility of limitations sources in classical contexts largely agrees with prior work, which has found that students are able to identify many limitations in simple introductory-level experiments~\cite{Sere1993}.

In contrast, we found that principles (e.g. the Heisenberg uncertainty principle) are the most consistently accessible sources of uncertainty across quantum contexts, both generically (73\% of students mentioned at least one principles source) 
and in the context of the single-slit experiment (58\% mentioned principles). We also found evidence that principles were accessible to many students even in the context of a quantum experiment in which principles were not particularly relevant. In the Stern-Gerlach experiment, the quantum mechanics at play predicts two discrete possible measurements, unlike the single-slit experiment, in which the quantum mechanical effects lead to a continuous distribution of possible measured values. In our Stern-Gerlach setup, moreover, only a single possible measurement was predicted by quantum mechanics, as we blocked one of the possible channels. In spite of this, 32\% of students drew on principles to explain the continuous distribution of measurements in our fictitious data distribution for the Stern-Gerlach experiment. Overall, our results indicate that principles are consistently accessible to students in quantum mechanics, even in contexts in which principles may not be relevant. 

This emphasis on principles in quantum mechanics largely agrees with our prior work, where most student interviewees identified principles sources of uncertainty in response to the same generic question about uncertainty in quantum mechanics and a similar single-slit scenario~\cite{Stein2019}. Baily and Finkelstein also found that students tended to draw on quantum principles to explain why multiple outcomes are possible in an experiment, particularly after taking a modern physics course~\cite{Baily2009a}.



\subsection{Limitations in quantum mechanics: Availability and accessibility}

Limitations seem to be \emph{available} in students' concept images of uncertainty in quantum mechanics, but their \emph{accessibility} may depend on multiple variables. 
Regarding availability, most students listed at least one source coded as limitations for both the single-slit (75\%) and Stern-Gerlach (81\%) experiments. This result differs from our previous findings, in which student interviewees almost exclusively discussed principles when considering sources of uncertainty in the single-slit experiment~\cite{Stein2019,Stump2020}. Key differences between these study formats suggest that additional question structure may have made limitations more \emph{accessible} to students in the current survey. 

In our earlier interview study, students were given a very brief description of the experiment and asked a variety of general questions about the shape of the data distribution. In the current study, the survey prompts followed a similar approach but with additional structure, such as a diagram of the experimental setup, a more detailed description of the equipment and procedures used in the experiment, and an explicit prompt to list as many causes of the distribution as students could think of. We infer that the students in our interviews likely only listed the \emph{most} accessible sources from their concept image (namely, principles). The added structure in our current study may have made a greater variety of sources accessible to students, including limitations in the experimental setup.

In addition, we find that the experimental quantum contexts generally make limitations sources much more accessible to students than with quantum mechanics outside a specific experimental context. In response to the generic question ``What comes to mind when you think about measurement uncertainty in quantum mechanics?'' only 19\% of students listed limitations sources. Unlike student responses to the experimental scenarios, this result agrees with our previous interview study~\cite{Stein2019,Stump2020}.
Without a specific experimental scenario to consider, and the additional scaffolding we provided in the survey experimental scenarios, students seem to be much less likely to draw on limitations in quantum mechanics.

Overall, our results suggest that limitations are \emph{available} as part of students' concept images of uncertainty in quantum mechanics. Students are able to access to limitations within highly structured experimental prompts, but access becomes limited without the structure or specific experimental context.

\subsection{Principles in classical mechanics: Availability and accessibility}

Principles may be \emph{available} in students' concept images of uncertainty in classical mechanics, with very limited \emph{accessibility}. Regarding availability, 56\% of students identified principles sources of uncertainty in the relevant Brownian motion experimental context.

Although principles were largely accessible to students in the context of the Brownian motion scenario, we found no evidence that principles were accessible to students when considering the projectile motion experiment, where only 6\% of students mentioned principles sources. On the one hand, the deterministic nature of the experiment justifies not mentioning principles of the physics theory as causing the spread in the data. On the other hand, students may reasonably discuss statistical principles as an important aspect of experimental measurement in the context of this experiment (e.g. multiple random measurements resulting in a normal distribution), as previous work has advocated teaching students this view of measurement~\cite{Evangelinos1999,Evangelinos2002,Allie2003,Pillay2008,Buffler2008}. Here, we saw no evidence that principles of measurement were available (let alone accessible) in students' concept images of uncertainty. This result may simply indicate that our experimental scenarios were ineffective at prompting access to this type of thinking or that statistical principles of measurement are not available to students. {We are also unsure whether \emph{experts} would list statistical principles of measurement for these scenarios. More research would be necessary to determine whether students or experts hold statistical principles in their concept images of uncertainty.

That very few students drew on principles for the generic classical question further supports the inaccessibility of principles in classical mechanics: only 7\% of students mentioned principles sources of uncertainty in response to this question.

Another possible explanation for the presence of principles sources in student responses to the Brownian motion experiment but not in the other classical-mechanics questions is that students may not view the Brownian motion experiment as being part of ``classical mechanics.'' Thus, students' responses to the Brownian motion scenario may not be representative of their concept images of uncertainty in classical mechanics. More research would be necessary to determine what range of physics contexts students see as part of classical mechanics or in which contexts principles are relevant sources of variability.

Overall, we observed that principles related to theoretical physics may be available to some degree in students' concept images of uncertainty in classical mechanics. These principles were only accessible, moreover, in a highly structured, non-deterministic classical experimental scenario but not in a deterministic classical experimental scenario.

\subsection{Summary and implications for future work}

We have observed that in the context of our survey questions, different knowledge pieces were accessible to students in classical and quantum mechanics: limitations in classical and principles in quantum. This result mirrors the epistemological split in student views about classical and quantum mechanics observed by Dreyfus \textit{et al.}~\cite{Dreyfus2019}. This split may be limiting for students' thinking about uncertainty, as limitations are relevant in quantum mechanics experimental applications~\cite{Fox2020} and principles of theoretical physics are relevant in stochastic classical systems, such as Brownian motion. Across contexts, students may also benefit from considering the statistical principles of measurement beyond the apparent limitations~\cite{Evangelinos1999,Evangelinos2002,Allie2003,Pillay2008,Buffler2008}. Even with the careful structure in our survey, however, students did not consistently draw on both limitations and  principles in all relevant contexts.

We also found evidence that may students draw on knowledge pieces irrelevant to the context at hand. For the Stern-Gerlach scenario, 32\% of students identified principles, primarily the Heisenberg uncertainty principle, even though the physics theory behind the experiment predicts a single-value outcome. Thus, the high accessibility of principles in quantum mechanics may be detrimental to students' reasoning about quantum-mechanical experiments.

These results are likely not surprising to physics educators and students. Quantum mechanics instruction tends to focus on toy models rather than practical applications and on mathematical calculations rather than conceptual reasoning~\cite{Johansson2018}. Thus, many students associate uncertainty in quantum mechanics with physical principles, such as the Heisenberg uncertainty principle, even when these principles are not relevant. 
The opposite is true in classical mechanics. In our survey, as in prior work~\cite{Sere1993,Holmes2015}, students most strongly associated uncertainty with limitations in lab experiments where the experimental aim is often to observe predicted behavior from a deterministic phenomenon. They simply may not associate uncertainty in classical mechanics with stochastic behavior, such as Brownian motion, that they learn about in a theory course context.

Our results prompt future research to identify which pieces of knowledge about sources of uncertainty experts believe are relevant in which contexts and how instruction can support students in recognizing those relevant knowledge pieces. This work also provides assessment questions complementary to procedural reasoning to explore student thinking about uncertainty in measurement, focusing on the conceptual and physical nature of uncertainty. In our subsequent work, we plan to use this characterization scheme to identify paradigms of students' conceptual thinking about uncertainty sources, which could then be used to evaluate the effects of instruction on students' thinking.

\acknowledgments{This material is based upon work supported by the National Science Foundation Graduate Research Fellowship under Grant No. DGE-2139899 and National Science Foundation Grants No.~DUE-1808945 and No.~DUE-1809178. We are grateful to our project evaluator, Ben Zwickl, for fruitful discussions of this work and to Peter Lepage, Courtney White, Andy Schang, and the Cornell Physics Education Research Lab for comments and feedback on this work over the last four years.}

\appendix

\section{Experimental scenarios}

\begin{figure*}
  \subfloat[Stern-Gerlach scenario]{\fbox{\includegraphics[width=.6\textwidth]{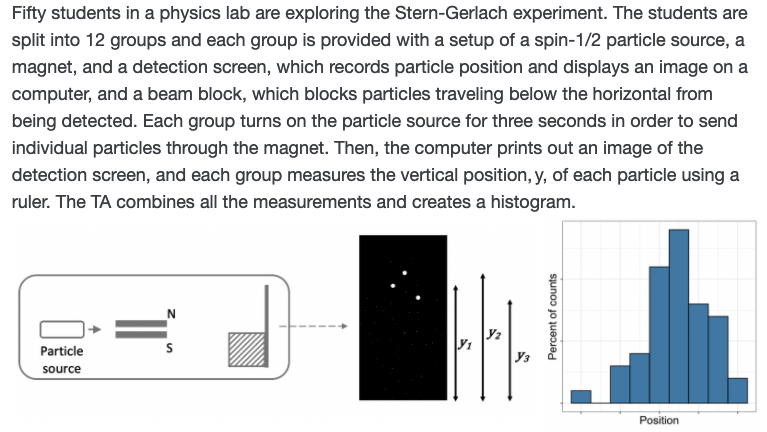}}\label{fig:SG}}\hspace{1em}
  \subfloat[Brownian motion scenario]{\fbox{\includegraphics[width=.6\textwidth]{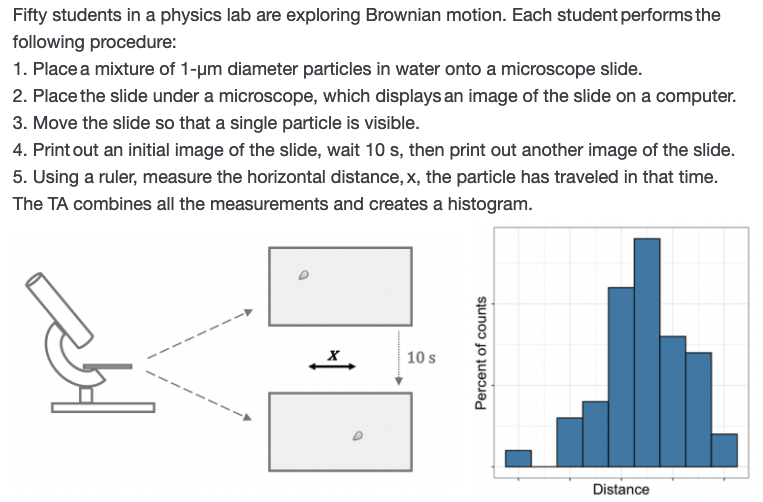}}\label{fig:BM}}\hspace{1em}
  \subfloat[Single-slit scenario]{\fbox{\includegraphics[width=.6\textwidth]{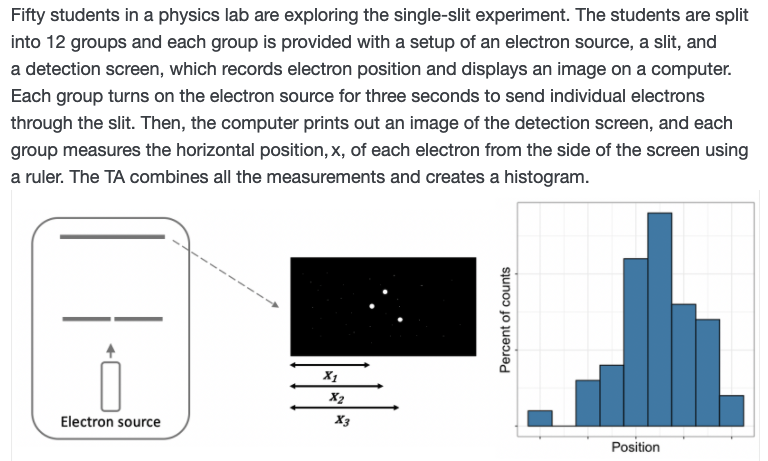}}\label{fig:SS}}
  \centering
    \caption{The descriptive text, diagram, and histogram of fictitious experimental data for the  upper-level experimental scenarios.}
    \label{fig:scenarios_full}
\end{figure*}

The full visual representations of the Stern-Gerlach, Brownian motion, and single-slit experimental scenarios in the survey are shown in Fig.~\ref{fig:scenarios_full}.

\section{Demographic information of study participants}

\begin{table}[tb] 
  \caption{Demographic information self-reported by the students included in this study ($N=150$). Students who marked two or more races are counted in each race category they chose.}
  \label{ta:demographics}
  \begin{ruledtabular}
    \begin{tabular}{lr}
    \textbf{Institution} & \\
    \quad California State University Fullerton & 8\\
    \quad Cornell University & 82\\
    \quad Michigan State University & 30\\
    \quad University of Colorado Boulder & 27\\
    \quad University of St. Andrews & 3\\
    \textbf{Year of college} \\
    \quad Second year (sophomore) & 9\\
    \quad Third year (junior) & 76\\
    \quad Fourth year + (senior) & 50\\
    \quad Graduate student & 11\\
    \quad Unspecified & 4\\
    \textbf{Gender}\\
    \quad Female & 40\\
    \quad Male & 104\\
    \quad Non-binary & 2\\
    \quad Unspecified & 4\\
    \textbf{Race/ethnicity}\\
    \quad American Indian or Alaska Native & 1\\
    \quad Asian or Asian American & 43 \\
    \quad Black or African American & 2 \\
    \quad Hispanic or Latinx & 18 \\
    \quad Native Hawaiian or other Pacific Islander & 1\\
    \quad Prefer to self-describe & 6\\
    \quad White & 78\\
    \quad Unspecified & 17\\
    \end{tabular}
  \end{ruledtabular}
\end{table}

Details about the institution, year, gender, and race/ethnicity of study participants are summarized in Table~\ref{ta:demographics}.

\section{Comfort and interaction between experimental scenarios and generic questions}

\begin{figure}
\includegraphics[width=.49\textwidth]{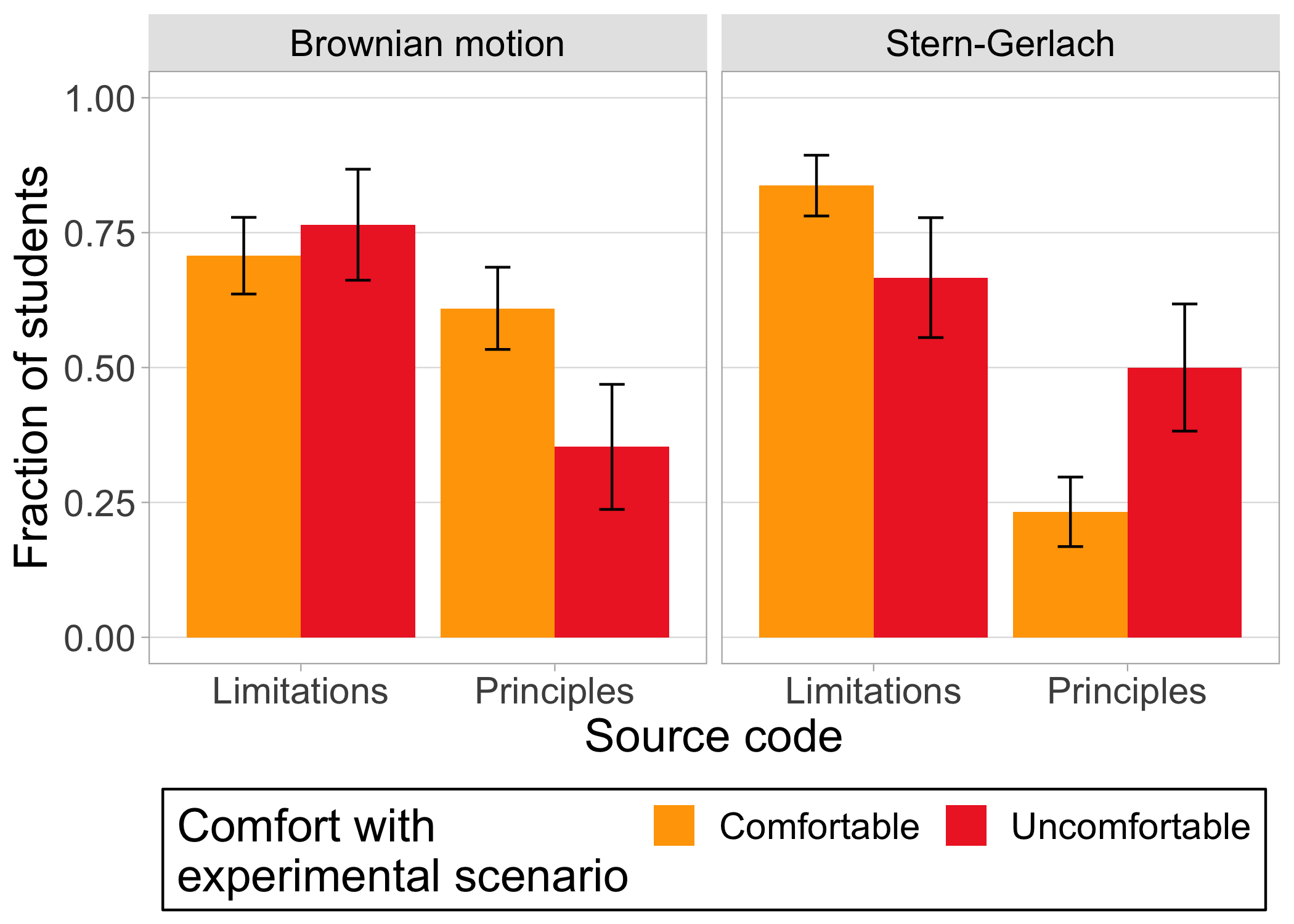}\\
  \centering
    \caption{Codes of sources listed by students with different comfort levels for the Brownian motion and Stern-Gerlach experiments, the two experiments with the lowest comfort levels. Error bars represent the standard error.}
\label{fig:BMSG_comfort}
\end{figure}

\begin{figure}
  \centering
  \includegraphics[width=.49\textwidth]{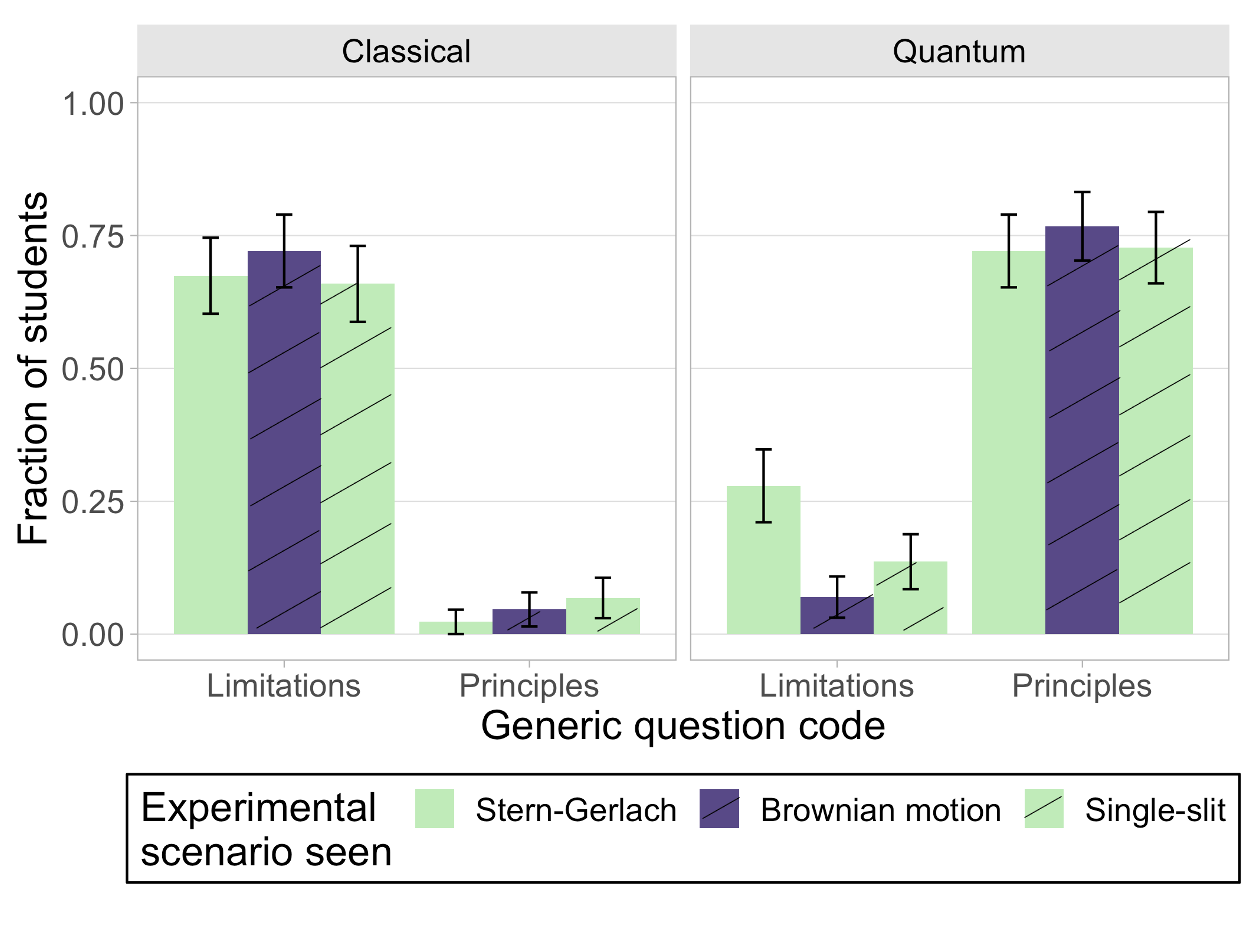}
  \caption{Codes for responses to the generic uncertainty questions, separated by what upper-level experiment each respondent saw during the survey. Physics paradigm is denoted by color (purple = classical, green = quantum), while theoretical expected outcome is denoted by pattern (solid = single value, striped = distribution). Error bars represent the standard error.}
  \label{fig:gen_exp}
\end{figure}

Fig.~\ref{fig:BMSG_comfort} shows the fraction of students who listed limitations and principles sources of uncertainty who were comfortable or uncomfortable with the Brownian motion and Stern-Gerlach experimental scenarios. Although we observe small differences in the fraction of comfortable and uncomfortable students who listed principles sources for each scenario, these differences are too small relative to their uncertainty to draw any strong conclusions.

Fig.~\ref{fig:gen_exp} shows the fraction of students who listed limitations and principles sources of uncertainty in response to the generic questions separated by what upper-level experiment each student saw. Overall, we observe no distinguishable differences in student responses to the generic questions based on experiment seen.

\section{Raw data tables}

The data corresponding to the main results figures (Fig.~\ref{fig:top_experiment} and Fig.~\ref{fig:top_gen}) are provided in Table~\ref{ta:top_experiment} and Table~\ref{ta:top_gen}. In each case, we include the total number of responses in each category, rather than percentages.

\begin{table}[tb] 
  \caption{Raw frequency data corresponding to Fig.~\ref{fig:top_experiment}.}
  \label{ta:top_experiment}
  \begin{ruledtabular}
    \begin{tabular}{llr}
    Source code & Experimental scenario & Count\\
    \hline
    Limitations & Projectile motion & 139\\
    Limitations & Stern-Gerlach & 48\\
    Limitations & Brownian motion & 44\\
    Limitations & Single-slit & 45\\
    Principles & Projectile motion & 9\\
    Principles & Stern-Gerlach & 19\\
    Principles & Brownian motion & 33\\
    Principles & Single-slit & 35\\
    \end{tabular}
  \end{ruledtabular}
\end{table}

\begin{table}[tb] 
  \caption{Raw frequency data corresponding to Fig.~\ref{fig:top_gen}.}
  \label{ta:top_gen}
  \begin{ruledtabular}
    \begin{tabular}{llr}
    Source code & Physics paradigm & Count\\
    \hline
    Limitations & Classical & 103\\
    Limitations & Quantum & 29\\
    Principles & Classical & 8 \\
    Principles & Quantum & 109\\
    \end{tabular}
  \end{ruledtabular}
\end{table}

\bibliography{_Sources_refs} 

\end{document}